\documentclass{pazha2}
\usepackage{epsf}
\usepackage{graphicx}
\usepackage{color}
\usepackage{multirow}
\definecolor{darkblue}{rgb}{0,0,0.9}
\def\a{$^{\mbox{\small a}}$}
\def\bb{$^{\mbox{\small b}}$}
\def\cc{$^{\mbox{\small c}}$}
\def\dd{$^{\mbox{\small d}}$}
\def\e{$^{\mbox{\small e}}$}
\def\f{$^{\mbox{\small f}}$}
\def\g{$^{\mbox{\small g}}$}

\begin{document}
\journalinfo{2017}{43}{10}{664}{736}{748}[676]
\sloppypar

\title{\Large Spectroscopic Study of the Optical Counterpart to the Fast X-ray
Transient IGR\,J17544-2619 Based on Observations at the 1.5-m RTT-150
Telescope}
\author{
  I. F. Bikmaev\address{1,2}\email{ibikmaev@yandex.ru},
  E. A. Nikolaeva\address{1,2}\email{ibikmaev@yandex.ru,
    evgeny.nikolaeva@gmail.com}, 
  V. V. Shimansky\address{1},
  A. I. Galeev\address{1,2},
  R.~Ya.~Zhuchkov\address{1,2},
  E.N. Irtuganov\address{1,2},
  S. S. Melnikov\address{1,2},
  N. A. Sakhibullin\address{1,2},
  S.A.Grebenev\address{3},
  and L.M. Sharipova\address{4}\\ [4mm]
$^1${\it Kazan (Volga Region) Federal University, Kazan, Russia}\\
\noindent
$^2${\it Academy of Sciences of Tatarstan, Kazan, Russia}\\
\noindent
$^3${\it Space Research Institute, Russian Academy of  Sciences, Moscow, Russia}\\
\noindent
$^4${\it Crimean Astrophysical Observatory, Russian Academy of
  Sciences, Nauchnyi, Crimea, Russia}}

\vspace{2mm}
\submitted{September 26, 2016}

\shortauthor{BIKMAEV et al.}
\shorttitle{SPECTROSCOPIC STUDY OF THE OPTICAL COUNTERPART TO IGR\,J17544-2619}

\begin{abstract}
\noindent
We present the results of our long-term photometric and
spectroscopic observations at the Russian-Turkish RTT-150
telescope for the optical counterpart to one of the best-known
sources, representatives of the class of fast X-ray transients,
IGR\,J17544-2619. Based on our optical data, we have determined
for the first time the orbital and physical parameters of the
binary system by the methods of Doppler spectroscopy. We have
computed theoretical spectra of the optical counterpart by
applying non-LTE corrections for selected lines and obtained the
parameters of the stellar atmosphere and the optical star
($T_{\rm eff}=33\,000$~K, $\log\,g=3.85$, $R=9.5\ R_{\odot},$
and $M=23\ M_{\odot}$). The latter suggest that the optical star
is not a supergiant as has been thought previously.\\

\noindent
{\bf DOI:} 10.1134/S1063773717100012

.
\end{abstract}


\section*{INTRODUCTION}
\noindent
The source IGR\,J17544-2619 belongs to supergiant
fast X-ray transients (SFXTs), a new population
of X-ray objects discovered by the INTEGRAL
observatory (Sunyaev et al. 2003b; in't Zand 2005;
Negueruela et al. 2006; Sguera et al. 2006; Grebenev
2009; Romano et al. 2014). X-ray binaries containing
an early-type (OB) supergiant and a neutron star
with a strong magnetic field (an X-ray pulsar) are
representatives of this population.
\begin{figure*}[thb]
\centering
\includegraphics[width=0.78\textwidth]{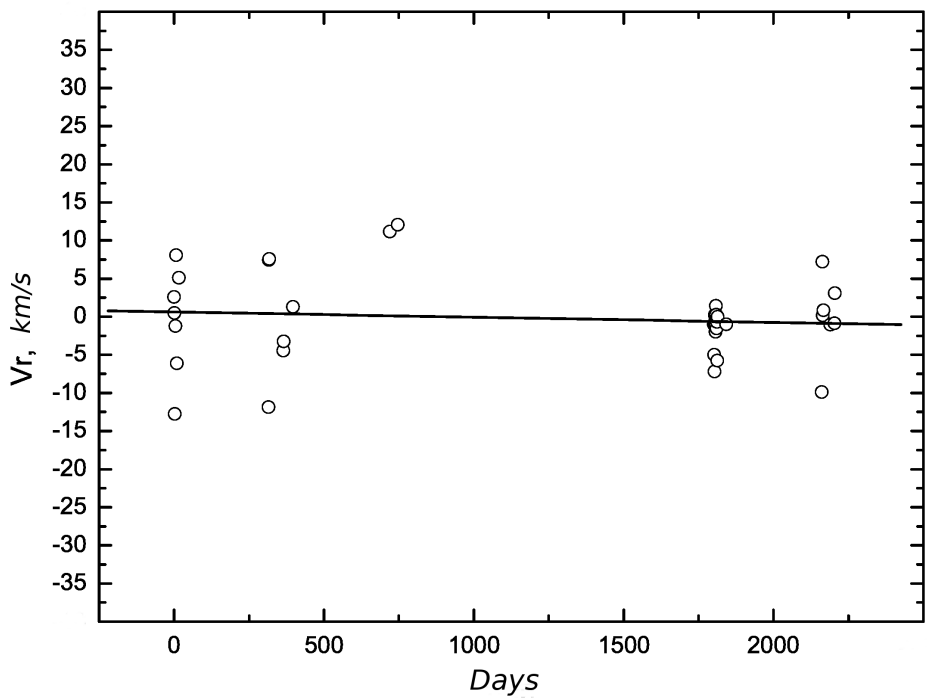}

\vspace{-6mm}

\caption{\rm Scatter of radial velocities from the interstellar
  NaD lines.}\label{Na_IGR17}

\end{figure*}

Supergiants are characterized by a strong outflow of material
(with a rate
$\sim10^{-6}-10^{-5}\ M_{\odot}\ \mbox{yr}^{-1}$). Such a dense
stellar wind must supply sufficient material to maintain
spherically symmetric accretion onto the neutron
star. Therefore, one would think that these binary systems must
be powerful X-ray sources with a luminosity
$\sim10^{37}\ \mbox{erg s}^{-1}$ (Grebenev and Sunyaev
2007). Actually, this is not the case.  Most of the time SFXTs
have a very low luminosity $\sim10^{32}-10^{33}\ \mbox{erg
  s}^{-1}$ and are not accessible for detection by wide-field
X-ray telescopes. However, they occasionally flare up for a
short ($\la1$ day) time, becoming the brightest objects in the
X-ray sky.  Their luminosity rises by 4--6 orders of magnitude
in a few minutes. Various models have been proposed to explain
the so unusual properties of SFXTs (the absence of persistent
emission, the brevity of outbursts, the huge range of luminosity
variations).  Historically the first model of a clumpy stellar
wind from a supergiant (in't Zand et al. 2004) did not explain
the observed wide range of luminosity variations, the model of a
highly asymmetric stellar wind (like the wind from Be stars,
Sidoli et al. 2007) did not explain the outburst brevity, and
the model of a magnetic barrier (Bozzo et al. 2008) suggested an
extremely strong neutron star magnetic field. The model by
Grebenev and Sunyaev (2007) (see also Bozzo et al. 2008), where
the accretion of matter is halted by a centrifugal barrier at
the magnetospheric boundary of a rapidly spinning neutron star
(the propeller effect), while the outbursts are explained by the
temporary overcoming of this barrier due to local fluctuations
of the stellar wind density and velocity, and the model by
Shakura et al. (2014), which suggests quasi-spherical subsonic
(settling) accretion onto a slowly spinning neutron star in a
steady state and an enhancement of accretion during outbursts
due to the development of Rayleigh-Taylor instability at the
magnetospheric boundary, remain topical and are widely
discussed. However, the latter model requires that the stellar
wind be strongly magnetized and the neutron star spin be too
slow ($P_s\ga1\,000$ s).
\begin{figure*}[t]
\centering
\framebox{\includegraphics[width=110mm]{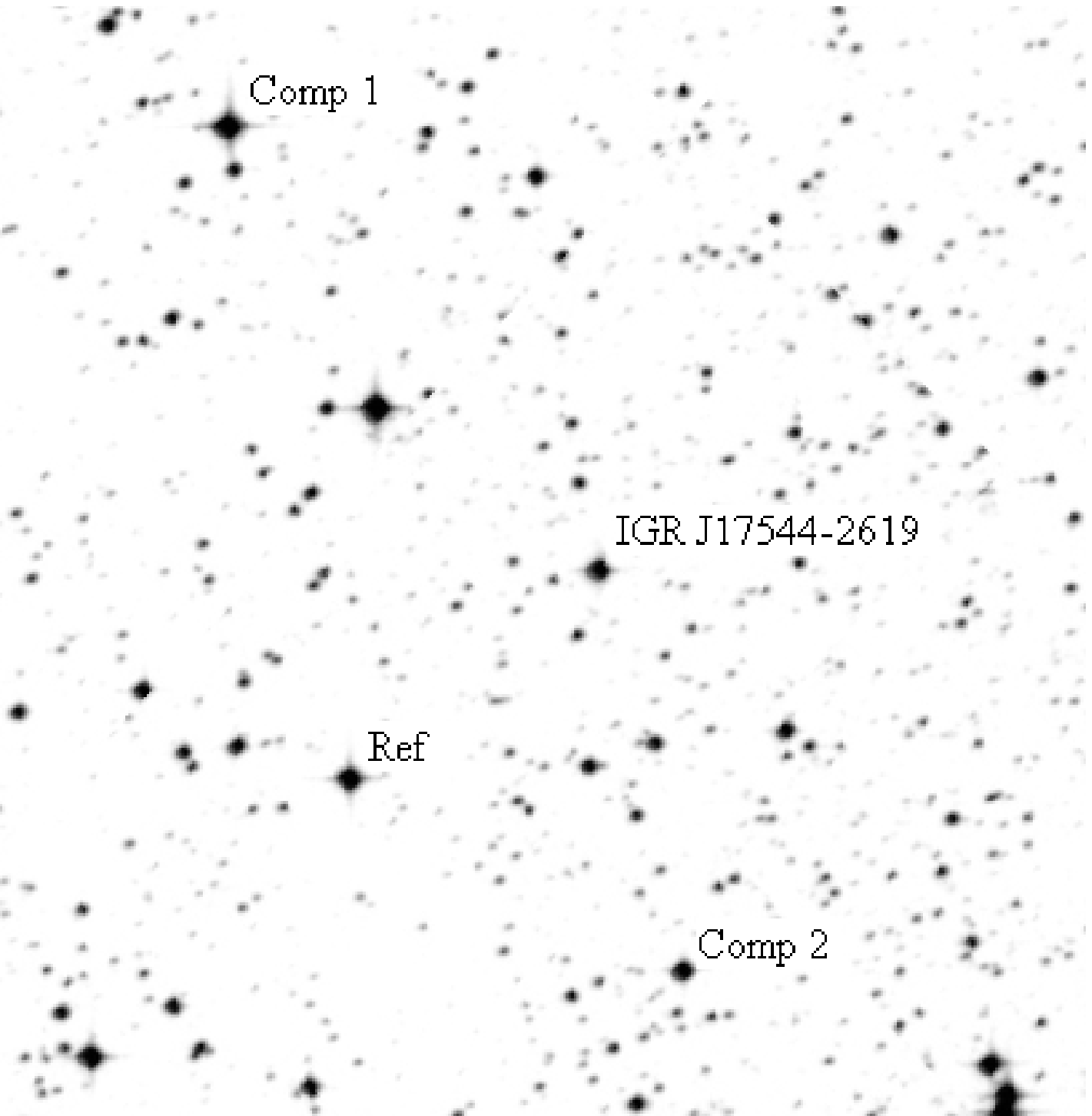}}
\caption{\rm Comparison stars Comp~1 (TYC 6849-1825-1) and
  Comp~2 ($\alpha=17\uh54\um22\fs578$,
  $\delta=-26\deg22\arcmin49\farcs43$), reference star Ref
  ($\alpha=17\uh54\um33\fs457$,
  $\delta=-26\deg21\arcmin23\farcs74$) and
  IGR\,J17544-2619.}\label{MapIGR17}
\end{figure*}

The source IGR\,J17544-2619 was discovered during an X-ray
outburst on September 17, 2003, (Sunyaev et al. 2003b) with the
IBIS/ISGRI gamma-ray telescope (Ubertini et al. 2003; Lebrun et
al. 2003) onboard the INTEGRAL observatory (Winkler et al.
2003). Recurrent outbursts (Grebenev et al. 2003, 2004) showed
that the source is similar to the other known transient
XTE\,J1739-302 (Sunyaev et al.  2003a; Smith et al. 2003, 2006),
from which such outbursts were observed, and allowed one to talk
about the discovery of a new population of X-ray sources,
SFXTs. The source IGR\,J17544-2619 became a canonical 
representative of this population; it was used to verify the
properties of SFXTs, to investigate their behavior, and to
construct their models. The properties of this transient are
indeed unique. For example, an outburst of IGR\,J17544-2619 with
a record luminosity $\sim 3 \times 10^{38}$ erg s$^{-1}$ was
detected on October 10, 2014, thereby extending the range of its
luminosity variations to six orders of magnitude (Romano et
al. 2015). Based on NuSTAR satellite observations of the source,
Bhalerao et al. (2015) have measured for the first time the
magnetic field strength of the neutron star in SFXTs, $\sim
1.5\times10^{12}$ G (from a cyclotron line $h\nu_c\simeq17$
keV); this value is typical for the neutron stars in X-ray
binaries. The optical counterpart to IGR\, J17544-2619 was
classified as an O9Ib supergiant with a mass of 25--28
$M_{\odot}$ (Pellizza et al. 2006) and a radius in the range
12.7 $R_{\odot} < R < 26.6 R_{\odot}$ (Rahoui et
al. 2008). Based on the minimum radius, using the previously
estimated masses of the supergiant, 25--28 $M_{\odot}$, and the
compact object, 1.4 $M_{\odot}$, and taking into account the
absence of regular outbursts, suggesting that there is no Roche
lobe overflow in the binary system, Clark et al. (2009) imposed
a constraint on the orbital eccentricity, $e \la 0.4$. Based on
archival INTEGRAL data and using the Lomb-Scargle method (Lomb
1976; Scargle et al. 1982), the authors also determined the
orbital period of the binary, $P_{b}\simeq (4.926 \pm 0.001)$
days. Subsequently, the period has been repeatedly confirmed
(for various ephemerides) on the basis of X-ray data:
$P_{b}=(4.9278 \pm 0.0002)$ суток, MJD 53\,732.632 (Drave et
al. 2012); $P_{b}=(4.9272 \pm 0.0004)$ days, MJD 55\,924.271
(Drave  et al. 2014); and $P_{b}=(4.92693 \pm 0.00036)$ days,
MJD 53\,732.65 (Smith 2014). However, as yet nobody has
determined it directly from optical data. In this paper we
present the results of our long-term observations of the optical
counterpart to this X-ray system at the Russian-Turkish RTT-150
telescope. Based on their analysis, we solved this problem.
\begin{table}[bh]
\begin{center}
{\small {\bf Table 1.}  $B$, $V$, $R$, and $I$ magnitudes of
  the\protect\\ standard and reference stars}
\end{center}

\begin{center}
\begin{tabular}{ccccc}
\hline
Star & $B$ & $V$ & $R$ & $I$ \\
\hline
SA\,104\underline{ }598 & 12.585 & 11.479 & 10.809 &  10.263 \\
Ref & 13.08 & 12.18 & 11.71 & 11.25 \\
\hline
\end{tabular}
\end{center}
\end{table}
\section*{OBSERVATIONS AND DATA ANALYSIS}
\noindent
The photometric observations of the object were begun with the
1.5-m Russian-Turkish RTT-150 telescope at the TUBITAK National
Observatory (Turkey) in 2003 shortly after the discovery of the
source. Unfortunately, although dozens of X-ray outbursts have
already been recorded from the source (see, e.g., Romano et
al. 2013), none of them was detected during the observations
being discussed.  From 2003 to 2006 an ANDOR ($2048 \times 2048$
pixels) CCD array was used for our photometry. Since 2007, after
the identification of the object with an optical star (Pellizza
et al. 2006), the photometric and spectroscopic observations
have been performed with the TFOSC instrument using a
medium-resolution ($2.5$ {\AA}) echelle mode. The duration of
each exposure was 30 min. The spectra taken during the night
were summed to increase the signal-to-noise ratio (S/N) in our
data. The mean S/N ratio of the summed spectrum was 70--120. The
spectra span the wavelength range $\lambda$\,4400--9150
{\AA}. The photometry was carried out in the $B$, $V$, $R$, $I$,
and white filters.
\begin{table*}[ptbh]
\begin{center}
{\small {\bf Table 2.} $B$, $V$, $R$, and $I$ magnitudes of IGR\,J17544-2619}
\end{center}

\begin{center}

\begin{tabular}{lcccc}
\hline
Data source& $B$ & $V$ & $R$ & $I$ \\
\hline
\hline
This paper (July 18, 2015) & $14.51 \pm 0.02$ & $12.77 \pm 0.01$ & $11.61 \pm 0.01$ & $10.35 \pm 0.01$ \\
This paper (June 30, 2016) & $14.52 \pm 0.01$ & $12.79 \pm 0.01$ & $11.63 \pm 0.01$ & $10.38 \pm 0.01$ \\
Pellizza et al. (2006) & $14.44 \pm 0.05$ & $12.65 \pm 0.05$ & $<11.9$ & $-$ \\
Zacharias et al. (2012) & $14.71 \pm 0.01$ & $12.94 \pm 0.01$ & $12.10 \pm 0.07$ & $-$ \\
DENIS database\a & $-$ & $-$ & $-$ & $10.38 \pm 0.01$ \\ \hline
\multicolumn{5}{l}{\a\ The DENIS database, the 3rd release, the CDS/ADC Collection of Electronic }\\
\multicolumn{5}{l}{\ \, Catalogues, 2263, 0 (2005).}\\
\end{tabular}
\end{center}

\begin{center}
{\small {\bf Table 3.} Radial velocities (km s$^{-1}$)}
\end{center}

\begin{center} 
\begin{tabular}{@{~}r@{~~}c@{~~}c@{~}c@{~}c|r@{~~}c@{~~}c@{~}c@{~}c@{~}}\hline
Date & HJD & Phase $\phi$ & $V_{r}$ & $\delta V_{r}$  &Date & HJD & Phase $\phi$ & $V_{r}$ & $\delta V_{r}$ \\
\hline
Aug. 1, 2007 & 2454314.3797 & 0.95910 & 22.8  & 2.7&July 18,
2012 & 2456127.3252 & 0.90505 & 25.8  & 6.5\\
Aug. 2, 2007 & 2454315.3842 & 0.16297 & -2.6  & 6.0&Aug. 16,
2012 & 2456156.2997 & 0.78557 & 32.0  & 3.6\\
Aug. 3, 2007 & 2454316.3704 & 0.36312 & -23.5 & 3.7&July 1,
2013 & 2456475.4444 & 0.55750 & 22.4  & 8.1\\ 
Aug. 5, 2007 & 2454318.3419 & 0.76325 & 29.3  & 3.4&July 3,
2013 & 2456477.4579 & 0.96616 & 24.2  & 3.0\\ 
Aug. 8, 2007 & 2454321.3142 & 0.36649 & 7.5   & 5.7&July 4,
2013 & 2456478.4011 & 0.15760 & -20.1 & 3.8\\ 
Aug. 17, 2007 & 2454330.3103 & 0.19231 & 0.3   & 1.8&July 6,
2013 & 2456480.4159 & 0.56651 & 17.5  & 4.9\\
June 11, 2008 & 2454629.3840 & 0.89073 & 37.4  & 2.7&July
29, 2013 & 2456503.3269 & 0.21640 & -17.1 & 5.2\\
June 12, 2008 & 2454630.5159 & 0.12045 & -10.4 &
4.8&Aug. 12, 2013 & 2456517.3178 & 0.05592 & 2.9   & 5.5\\
June 13, 2008 & 2454631.4483 & 0.30969 & 6.8   &
3.7&Aug. 14, 2013 & 2456519.3269 & 0.46367 & 6.6   & 1.8\\
July 30, 2008 & 2454678.3095 & 0.82040 & 33.4  & 4.7&June
23, 2014 & 2456832.3755 & 0.99839 & 7.8   & 4.0\\
July 31, 2008 & 2454679.2955 & 0.02051 & 2.9   & 2.6&June
24, 2014 & 2456833.4112 & 0.20858 & -5.5  & 3.9\\
Aug. 31, 2008 & 2454710.2850 & 0.30997 & -8.0  & 4.8&June
26, 2014 & 2456835.4433 & 0.62102 & 21.6  & 3.3\\
July 21, 2009 & 2455034.3843 & 0.08747 & -7.1  & 4.9&June
27, 2014 & 2456836.4180 & 0.81883 & 34.1  & 6.6\\
Aug. 16, 2009 & 2455060.3641 & 0.36020 & -2.4  & 4.2&June
28, 2014 & 2456837.4328 & 0.02479 & -0.9  & 3.5\\
July 6, 2012 & 2456115.3995 & 0.48467 & 9.2   & 3.6&July 1,
2014 & 2456840.3763 & 0.62218 & 7.0   & 1.8\\
July 7, 2012 & 2456116.4460 & 0.69706 & 32.2  & 5.5&July 2,
2014 & 2456841.4278 & 0.83560 & 24.0  & 1.4\\ 
July 8, 2012 & 2456117.3731 & 0.88523 & 42.3  & 5.3&July 24,
2014 & 2456863.3582 & 0.28648 & -19.2 & 5.0\\
July 9, 2012 & 2456118.3721 & 0.08799 & -3.0  & 4.5&July 26,
2014 & 2456865.3656 & 0.69387 & 23.8  & 0.6\\
July 10, 2012 & 2456119.3389 & 0.28420 & 4.8   & 3.4&July
28, 2014 & 2456867.3430 & 0.09521 & -26.2 & 2.7\\
July 11, 2012 & 2456120.3559 & 0.49061 & 24.5  & 5.2&July
21, 2015 & 2457225.3542 & 0.75528 & 26.1  & 5.6\\
July 12, 2012 & 2456121.3764 & 0.69772 & 38.7  & 7.1&June
30, 2016 & 2457570.4555 & 0.79525 & 24.8  & 0.9\\
July 13, 2012 & 2456122.3747 & 0.90032 & 28.5  & 1.7&July 1,
2016 & 2457571.4399 & 0.99503 & -4.5  & 7.6\\
July 14, 2012 & 2456123.3463 & 0.09753 & -5.7  & 5.0&July 2,
2016 & 2457572.4317 & 0.19633 & -13.7 & 4.4\\
July 15, 2012 & 2456124.4007 & 0.31152 & -6.9  & 6.0&July 3,
2016 & 2457573.4164 & 0.39617 & -13.6 & 2.6\\
July 16, 2012 & 2456125.3193 & 0.49794 & 31.1  & 6.7&July 6,
2016 & 2457576.3952 & 0.00073 & -14.5 & 7.2\\
\hline
\end{tabular}
\end{center}


\begin{center} 
{\small {\bf Table 4.} Orbital parameters of IGR\,J17544-2619\a}
\end{center}

\begin{center} 
\begin{tabular}{@{}c|r@{$\pm$}l@{}}
\hline
Parameters &\multicolumn{2}{c}{Values \makebox[1cm]{}}\\
\hline
 \makebox[2cm]{} $P_b$\bb\  \makebox[2cm]{}& 4.927206 & 0.0002 \\
  $T$\cc\ & 54314.6 & 0.1  \\
  $e$\dd\ & 0.42 & 0.07  \\
  $\omega$\e\ & 271 & 12  \\
  $K$\f\ & 21 & 2  \\
  $\gamma$\g\ & 9 & 1  \\
\hline

\multicolumn{3}{l}{}\\ [-3mm]
\multicolumn{3}{l}{\a\ Found using A.A. Tokovinin's {\it ORBIT\/} code.}\\
\multicolumn{3}{l}{\bb\ The orbital period, days.}\\
\multicolumn{3}{l}{\cc\ The time of periastron passage (JD-2\,400\,000), days.}\\
\multicolumn{3}{l}{\dd\ The orbital eccentricity.}\\
\multicolumn{3}{l}{\e\ The longitude of periastron in degrees.}\\
\multicolumn{3}{l}{\f\ The semi-amplitude of the radial velocity curve, km s$^{-1}$.}\\
\multicolumn{3}{l}{\g\  The center-of-mass velocity, km s$^{-1}$.}\\
 \end{tabular}
\end{center}

\end{table*}
\begin{figure*}[th]
  \centering
  \vspace{-6mm}
	\includegraphics[width=0.78\textwidth]{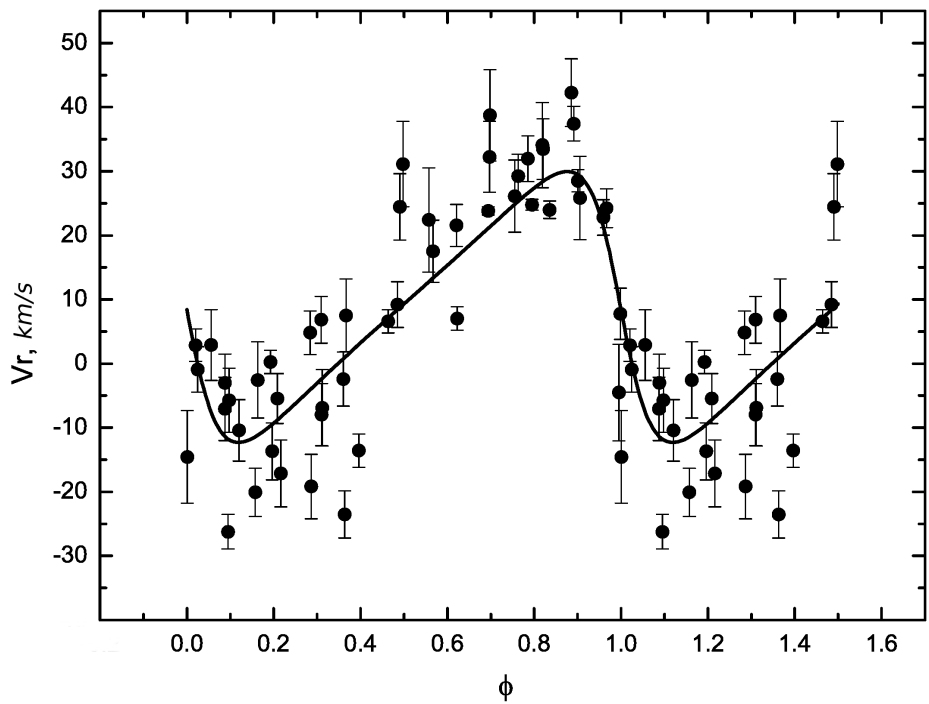}
        \vspace{-6mm}
        \caption{\rm Radial velocity curve of the optical counterpart to
  IGR\,J17544-2619.}\label{Vr_IGR17}

\end{figure*}

The reduction of the spectra was performed in
the {\it DECH\/} software package (Galazutdinov 1992, 2007) and
consisted of the following standard operations: bias
subtraction, spectrum extraction, cosmic-ray particle and
defective pixel removal, continuum placement, wavelength
calibration, and spectral line radial velocity measurements. The
radial velocities were measured by the cross-correlation method
from the He\,I $\lambda$\,5876, 6678, and 7065 {\AA} lines,
because the hydrogen line profiles are severely distorted by the
stellar wind from an O star. We controlled the instrumental
accuracy of our measurements based on the radial velocity
measurements of the interstellar sodium lines (Na\,I
$\lambda$\,5896, 5890 {\AA}). The rms error in the radial
velocities of the Na lines was $6\ \mbox{km s}^{-1}$
(Fig.\,\ref{Na_IGR17}) and was subsequently used as the
instrumental uncertainty.

\section*{RESULTS}
\subsection*{Photometry}
\noindent
The magnitude of the optical star being investigated was
measured relative to the reference star Ref
(Fig.\,\ref{MapIGR17}); whose magnitude was measured using the
Landolt photometric standard SA\,104\underline{ }598 (Table\,1).
No photometric variability of the object was detected at an
accuracy level of $\pm 0\fm02$ controlled with using the comparison
stars Comp1 and Comp2. This is evidenced by the long series of
their observations performed in the white filter. The measured $B$,
$V$, $R$, and $I$ magnitudes on different dates of observations
(July 18, 2015, and June 30, 2016) also lie within the error
limits and are in good agreement with the previously published
data (Table\,2).
\begin{figure*}[th]
\centering
\begin{minipage}[t]{0.48\textwidth}
  \hspace{-4.5mm}\includegraphics[width=1.166\textwidth]{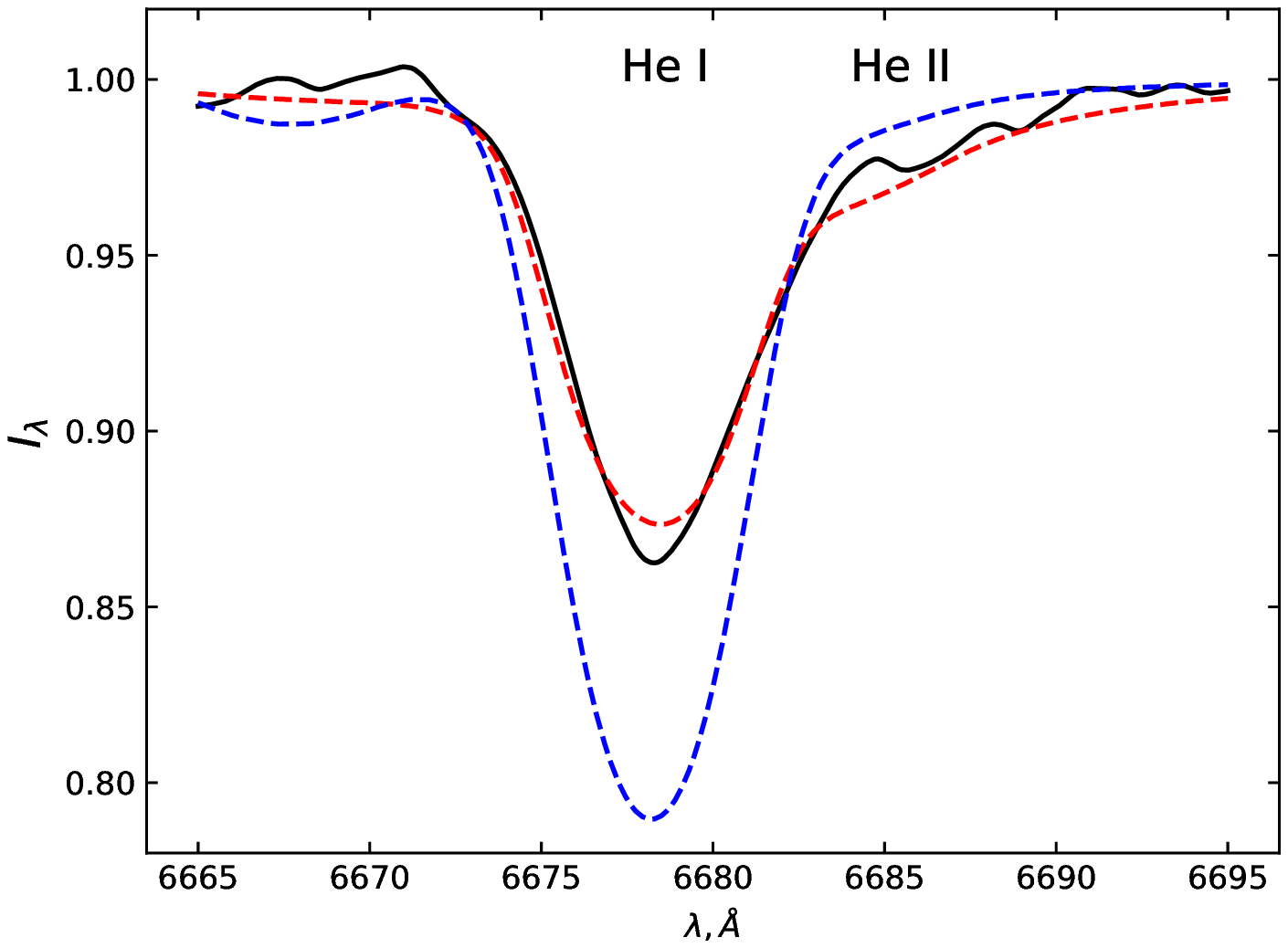}
\end{minipage} ~~ \begin{minipage}[t]{0.48\textwidth}
  \hspace{-4.5mm}\includegraphics[width=1.166\textwidth]{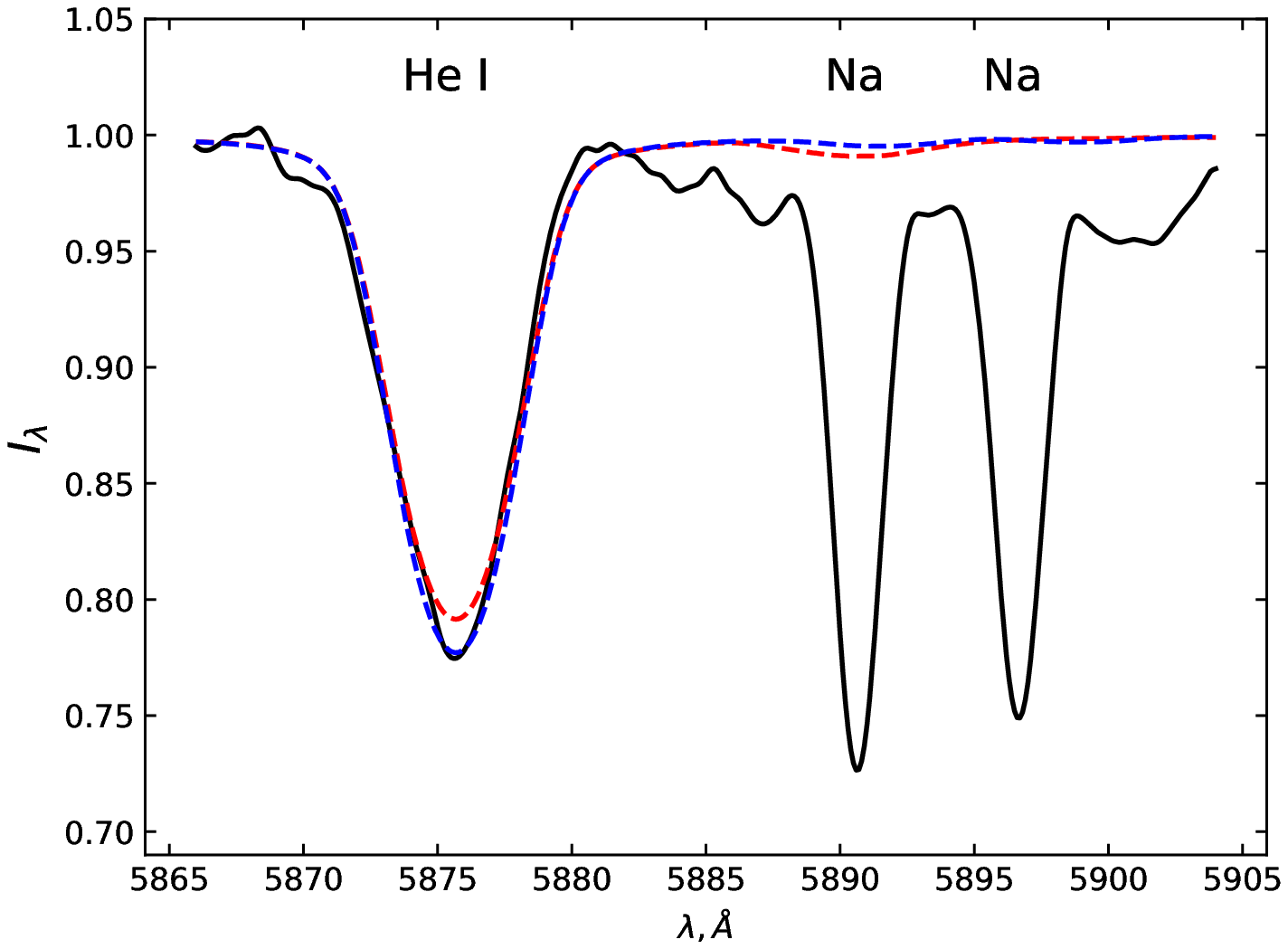} 
\end{minipage}

\vspace{-2mm}

\caption{\rm Observed and theoretical He\,I $\lambda$\,6678
  {\AA} and He\,II $\lambda$\,6683 {\AA} line profiles. The
  observed spectrum is indicated by the black solid line; the
  red dashed line is the theoretical spectrum that is in best
  agreement with the observed one, $T_{\rm eff}=33\,000$ K and
  $\log\,g=3.85$; the blue dashed line is the synthetic spectrum
  typical for a supergiant with $T_{\rm eff}=29\,000$ K,
  $\log\,g=3.20$. The distribution of colors on the succeeding
  graphs is analogous.}\label{He6678}

\vspace{-2mm}

\caption{\rm Observed and theoretical profiles of the He\,I
  $\lambda$\,5875 {\AA} line and the interstellar NaD lines
  demonstrating the spectral resolution.}\label{He5875}
\end{figure*}

\subsection*{Determining Orbital Parameters of the Binary System}
\noindent
Based on the spectra taken in 2007--2011, we measured the radial
velocities of the optical counterpart to IGR\,J17544-2619
(Table\,3) and then made an attempt to find the orbital
period of the binary system. The derived preliminary period,
12.18 days (Nikolaeva et al. 2013), contradicted the period
found by Clark et al. (2009) from X-ray data, $P_b\simeq (4.926
\pm 0.001)$ days. Therefore, the observations were continued,
and in 2012 we obtained a continuous 13-day series of spectra
whose analysis gave a period consistent with Clark's
estimate. Followup observations allowed the period to be
determined with an even higher accuracy. The orbital period was
determined by the Lafler-Kinman (1965) and Deeming (1975)
methods using V.P. Goransky's {\it WinEfk\/} code from the
spectroscopic data spanning the period of observations
2007--2016. The tables of possible periods can be found in
Nikolaeva et al. (2017). The orbital parameters of the binary
system were determined from the best period found using
A.A. Tokovinin's {\it ORBIT\/} code by the gradient-descent
method (Table\,4). The derived radial velocity curve is
presented in Fig.\,\ref{Vr_IGR17}. Given the period and the
semi-amplitude of the radial velocity curve, we calculated the
mass function of the binary system $f(m) =
(3.6\pm1.0)\times10^{-3} M_{\odot}$ and the line-of-sight
projection of the semimajor axis of the optical counterpart
$a_{v} \sin i = (0.00871\pm 0.00086)\ \mbox{AU}.$

\subsection*{Determining the Mass of the Optical Counterpart}
\noindent
The first steps in determining the mass of the optical star of
IGR\,J17544-2619 were taken in Pellizza et al. (2006), where the
star was found from the ratio of the He\,II $\lambda$\,4541
{\AA}/He\,I $\lambda$ 4471 {\AA} line intensities to be of
spectral type O9, while its luminosity class Ib was determined
based on the presence of the He\,II $\lambda$\,4686\,{\AA} line,
which is a criterion for this luminosity class (the spectral
atlas of Walborn and Fitzpatrick 1990). Thus, the mass of the
optical counterpart was limited by the range 25--28 $M_{\odot}$.

Gimenez-Garcia et al. (2016) estimated the mass of the
supergiant, $(25.9\pm 1.0)\ M_{\odot}$, by computing its
synthetic spectra using the Potsdam Wolf-Rayet model atmosphere
code. The effective temperature was calculated from the ratios
of the He\,I\,/\,He\,II and Si\,III\,/\,Si\,IV lines, while
$\log\,g_{\rm eff}$ was derived from the broadened wings of the
Balmer lines and He\,II lines. The stellar radius $R = 17
R_{\odot}$ was determined from the luminosity and the effective
temperature found from modeling; the mass of the optical
counterpart was then calculated using $\log\,g_{\rm eff}$.
 
In our paper we independently determined the physical parameters
of the optical counterpart by studying the profiles of spectral
lines forming in its atmosphere.  For our analysis we selected
the observed spectrum containing a maximum set of absorption
lines in a wide wavelength range, $\lambda$\,5400--7080 {\AA},
that had no appreciably contribution from the emission
components, i.e., were undistorted by the wind outflows from the
star. We computed theoretical normalized spectra for the optical
counterpart to IGR\,J17544-2619 using the {\it STAR\/} code
(Menzhevitski et al. 2014), which takes into account the
blanketing in atomic and molecular lines and the non-LTE effects
for selected atoms and ions (in our case, H\,I, He\,I, He\,II,
C\,II, C\,III, and Mg\,II). Hydrostatic, plane-parallel model
stellar atmospheres with specified sets of parameters $T_{\rm
  eff},\ \log\,g,$ [He/H], and [M/H] were computed using the
{\it ATLAS12\/} software (Castelli and Kurucz 2004) adapted to personal
computers and kindly provided to us by V. Tsymbal.  In the
initial model computations we specified a solar chemical
composition according to the data from Anders and Grevesse
(1989), while in the final ones it was changed based on the
abundance determinations for a number of elements when analyzing
the observed spectrum (see below).
\begin{figure*}[th]
\centering
\begin{minipage}[t]{0.48\textwidth}
  \hspace{-4.5mm}\includegraphics[width=1.166\textwidth]{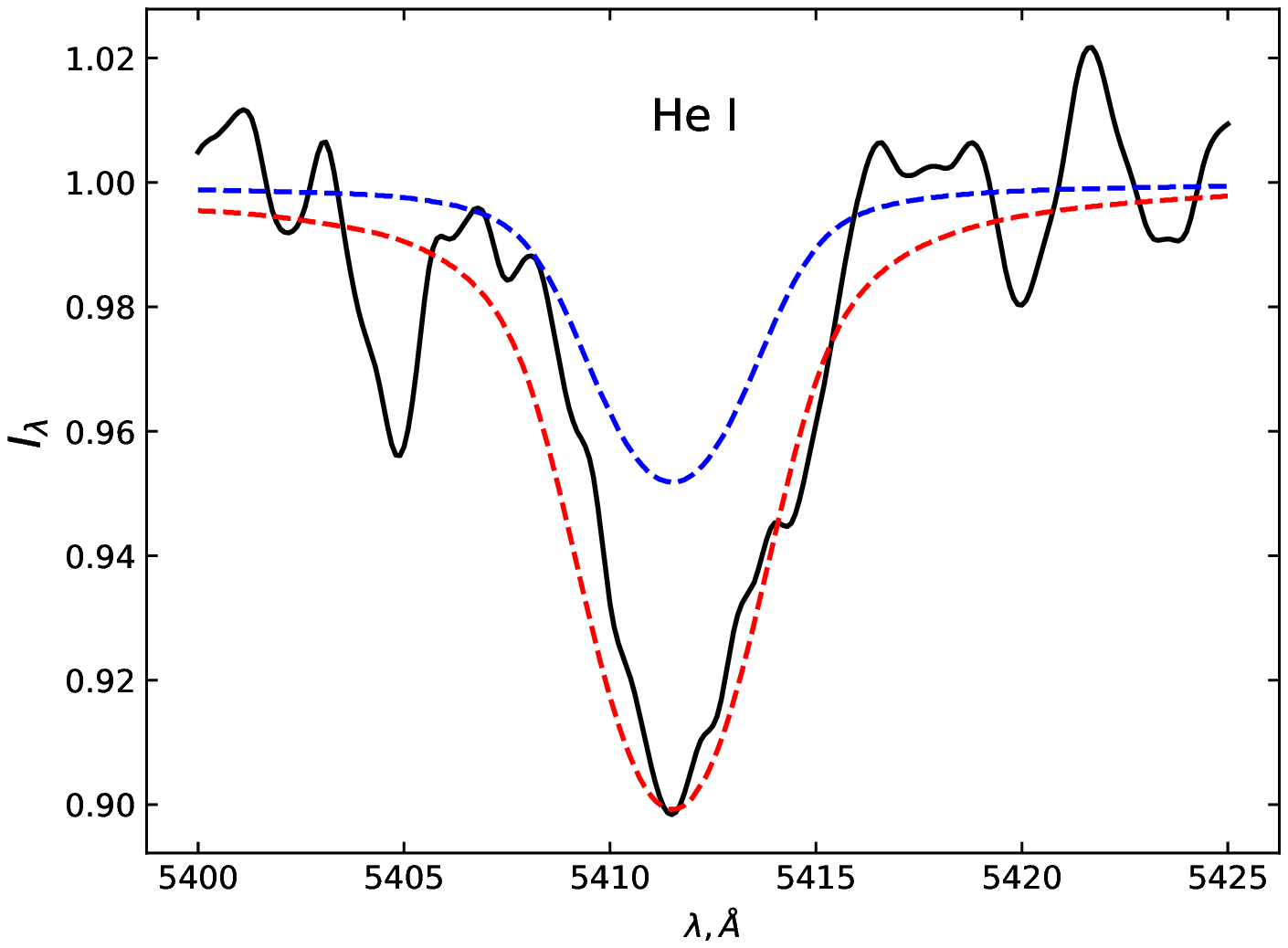}

  \vspace{-3mm}

  \caption{\rm Observed and theoretical He\,II $\lambda$\,5411
    {\AA} line profiles.}\label{He5411}
\end{minipage} ~~ \begin{minipage}[t]{0.48\textwidth}
  \hspace{-4.5mm}\includegraphics[width=1.166\textwidth]{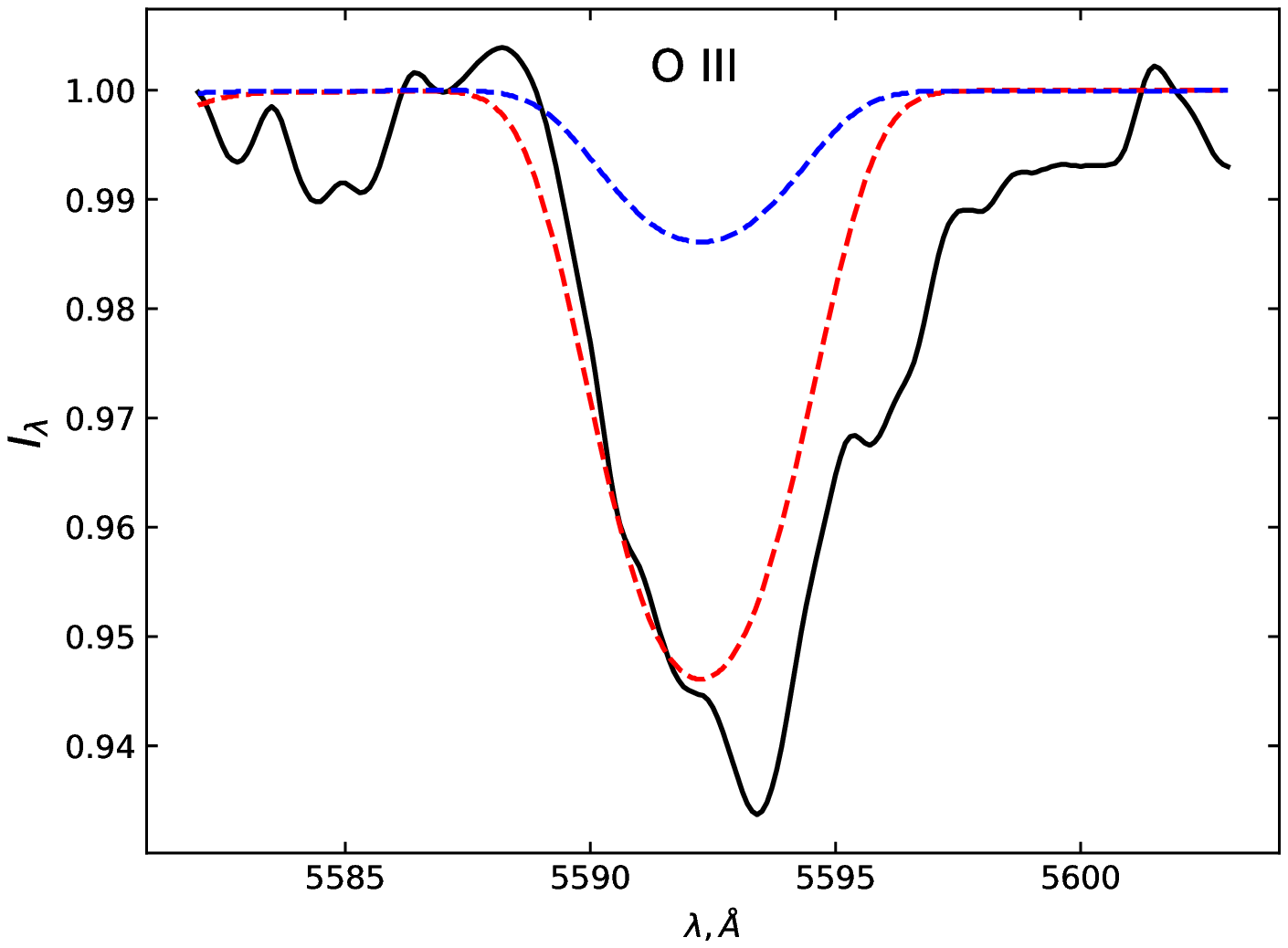}

  \vspace{-3mm}

  \caption{\rm Observed and theoretical O\,III $\lambda$\,5592
    {\AA} line profiles.}\label{O5592}
\end{minipage}
\end{figure*}

Based on the constructed models, when solving the transfer
equation by Hermitte's method, we calculated the specific
intensities of the optical emission emerging at three fixed
angles $\theta^{\prime}$ relative to the normal to the stellar
surface. We took into account all of the continuous absorption
sources tabulated in $STARDISK$ (Suleymanov 1992) and $SPECTR$
(Shimansky et al. 2012) and $\sim $2\,000\,000 lines from the
lists by Kurucz (1994) and Castelli and Kurucz (2004). In our
computations of the H\,I and He\,II line profiles we used the
Vidal--Cooper--Smith (1973) and Griem (1960) theories. The
profiles of the remaining lines were modeled by taking into
account the thermal motion of atoms and the microturbulence with
velocity $\xi_{\rm turb}$, the natural damping, the Stark
broadening according to the data from Kurucz and Furenlid (1979)
and the van der Waals broadening according to Gray's formula
(Unsold 1955). The departures from LTE in the stellar
atmospheres were found by applying 22-level H\,I, 38-level
He\,I, 23-level He\,II, 50-level C\,II, 94-level C\,III,
45-level Mg\,II models and the {\it NONLTE\/} software package
(Sakhibullin 1983) in the form of Menzel coefficients for the
energy levels of a selected atom and were passed to the {\it
  STAR\/} code according to the technique developed by Shimansky
(2012). The basic solar chemical composition was specified in
accordance with the data from Anders and Grevesse (1989). Then,
the stellar surface was divided into fields with a $2^{\circ}$
step in two orientation angles. For each field we found the
area, the observer's visibility angle $\theta$, and the
line-of-sight component of the rotational velocity.  The stellar
spectrum was computed by summing the intensities of the fields
interpolated to the angle $\theta$ and shifted in wavelength in
accordance with their radial velocity. The final spectrum was
convolved with the response function of the spectrograph
specified by a Gaussian with a FWHM corresponding to the
spectral resolution.

A proper analysis of the observed spectrum with the
determination of the binary parameters is possible under the
condition of its normalization based on its comparison with the
theoretical spectrum. Therefore, the renormalization was
performed so as to accurately match the mean residual
intensities in the observed and theoretical spectra in the
segments that are definitely free from lines. The positions of
these segments were chosen by analyzing the theoretical spectrum
provided that the intensity exceeded $I_{\lambda} = 0.98$ and
that there were no unknown lines in the observed spectrum at a
given wavelength. The observed spectrum contains a complex
emission-absorption $H_{\alpha}$ profile, the He\,I
$\lambda$\,5875, 6678, 7065 {\AA}, He\,II $\lambda$\,5411, 6406,
6526, 6683, 6890 {\AA}, C\,III $\lambda$\,5826 {\AA}, C\,IV
$\lambda$\,5801, 5812 {\AA}, O\,III $\lambda$\,5508, 5592 {\AA}
absorption lines, and the C\,III $\lambda$ 5694, 5696 {\AA}
emission blend. Obviously, the $H_{\alpha}$ profile is formed
almost completely in the stellar wind and is unsuitable for
analyzing the stellar characteristics. The remaining lines are
formed in the stellar atmosphere and can be used to determine
its parameters. The presence of helium and carbon lines in two
successive ionization stages in the spectra theoretically allows
us to investigate their ionization equilibria and to find the
abundances of elements in the atmosphere of the star, its
effective temperature $T_{\rm eff}$, and surface gravity
$\log\,g$ in a joint analysis. Note that the neglect of
departures from LTE for C\,IV in our technique and the weakness
of the C\,III and C\,IV lines can cause significant errors in
constructing the C\,III/C\,IV equilibrium. Therefore, when
analyzing the spectrum and determining the parameters of
IGR\,J17544-2619, we focused our attention on a proper
description of the observed He\,I and He\,II line profiles,
while the C\,III and C\,IV lines were used to find the carbon
abundance and to check the results. However, it should be noted
that the presence of C\,IV and O\,III lines in the stellar
spectra unambiguously points to its effective temperature
appreciably exceeding $T_{\rm eff}=30\,000$\,K.
\begin{figure}[th]
\vspace{-6mm}
  
\hspace{-4.4mm}\includegraphics[width=0.565\textwidth]{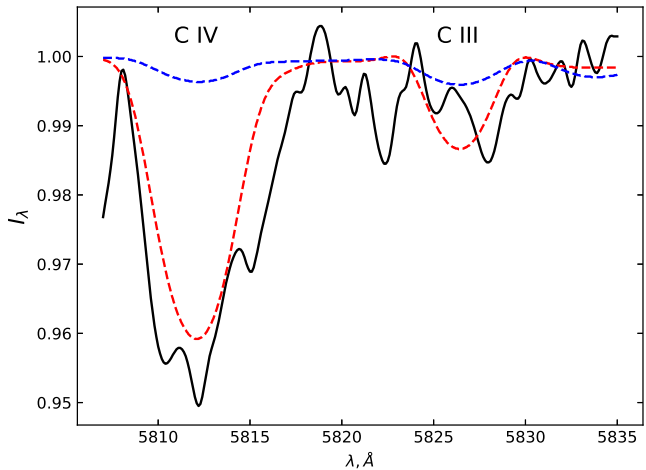}

\vspace{-3mm}

\caption{\rm Observed and theoretical C\,IV $\lambda$\,5812
  {\AA} and C\,III $\lambda$\,5826\,{\AA} line profiles.}\label{C5812} 
\end{figure}

The parameters and chemical composition of IGR\,J17544-2619 were
obtained by the method of successive approximations until an
optimal match between the observed and theoretical line profiles
was achieved. The effective temperature was determined by the
requirement that the unblended He\,I $\lambda$\,5875, 6678 {\AA}
lines and the strongest unblended He\,II $\lambda$\,5411 {\AA}
line be described simultaneously.  The surface gravity was
estimated by analyzing the profiles of the He\,I
$\lambda$\,5875, 6678 {\AA} lines with different sensitivities
to the adopted $\log\,g$. To find the microturbulent velocity $\xi_t$,
we used the O\,III $\lambda$\,5508, 5592 {\AA} and C\,IV
$\lambda$\,5801, 5812 {\AA} lines of different intensities. This
approach allowed, on the whole, satisfactory agreement between
the observed and theoretical spectra of the object to be
achieved at the follo
wing parameters: $T_{\rm eff}=(33\,000 \pm
1\,000)$\, K, $\log\,g=(3.85 \pm 0.15)$, [He/H] = 0.45 dex,
[C/H]= 0.4 dex, [N/H] = 0.5 dex, [O/H] = 0.8 dex, $\xi_t=17$\ km
s$^{-1}$.

Figures \ref{He6678}--\ref{C5812} present the deepest He\,I
$\lambda$\,6678 {\AA} and He\,II $\lambda$\,6683 {\AA}, He\,I
$\lambda$\,5875 {\AA}, He\,II $\lambda$\,5411 {\AA}, O\,III
$\lambda$\,5592 {\AA}, C\,IV $\lambda$\,5812 {\AA}, and C\,III
$\lambda$\,5826 {\AA} lines. The computed theoretical spectrum is
presented in comparison with the observed spectrum and the
theoretical one for atmospheric parameters typical for a
supergiant, $T_{\rm eff}=29\,000$\, K, $\log\,g=3.20$\ dex, and
$\xi_t=25$\,km\,s$^{-1}$, and determined in Gimenez-Garcia
et al. (2016). Obviously, using the previously published
parameters does not allow a number of features in the observed
spectrum, in particular, the O\,III, C\,IV, He\,I
$\lambda$\,6678 {\AA}, He\,II $\lambda$\,5411 {\AA} lines and a
number of weaker He\,II lines, to be described. Therefore, we
believe the set of stellar atmosphere parameters we found to be
more reliable and to admit a proper match of the observed and
theoretical spectra. Note that the enhanced abundances of helium
and light elements we found\footnote{Note that the accuracy of
  the helium abundance determination when analyzing the lines of
  the He\,I and He\,II ionization stages with a complete
  allowance for the departures from LTE is about 0.15 dex. The C
  abundance derived from the lines of two ionization stages (one
  in non-LTE) has errors of about 0.22 dex; the O abundance has
  an error up to 0.25 dex.  Therefore, on the whole, the C and O
  overabundances we found may be deemed real.} can be associated
with the complex evolution of IGR\,J17544-2619: these elements
could be synthesized in the interiors of the primary component
after its transformation into a supergiant, i.e., when the
binary system was passing the common-envelope phase. The partial
transfer of matter to the secondary component (the now observed
optical star) occurring at this time could contribute to its
enrichment by newly synthesized elements. Note that similar
overabundances of light elements have previously been found in
other high-mass binaries with relativistic components (for
example, in V1357 Cyg, see Shimansky et al. 2012).

\begin{figure*}[th]
  \centering

  \vspace{-3mm}
  
\includegraphics[width=130mm]{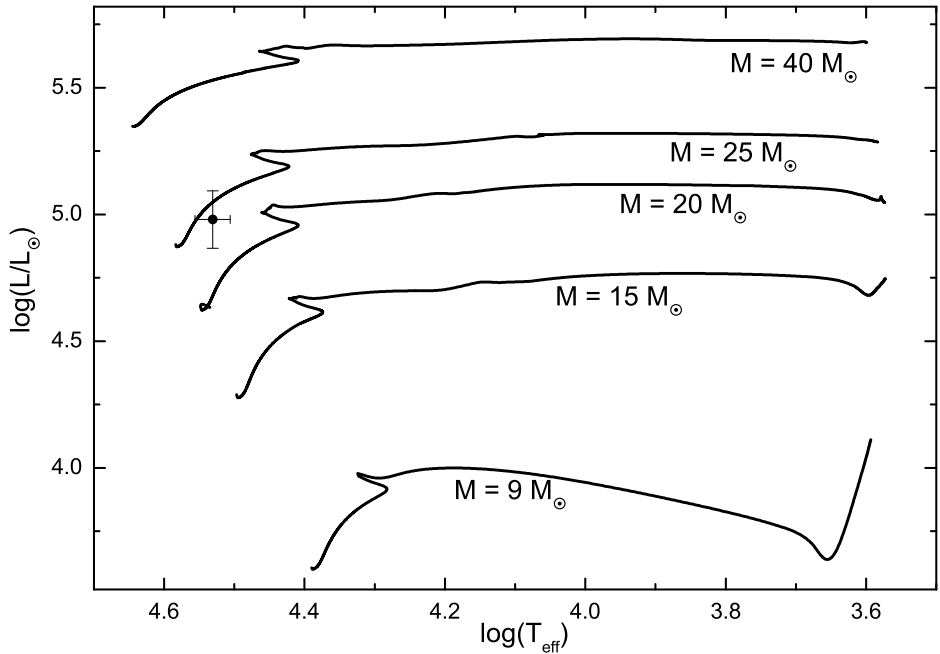}
\caption{\rm Position of the optical counterpart to
  IGR\,J17544-2619 on the diagram $\log T_{\rm
    eff}-\log(L/L_{\odot})$ in comparison with the evolutionary
  tracks.}\label{Trek_IGR17}

\vspace{-2mm}

\end{figure*}

Figure \ref{Trek_IGR17} shows the position of the optical
counterpart to IGR\,J17544-2619 on the $\log T_{\rm
  eff}-\log(L/L_{\odot})$ diagram in comparison with the
evolutionary tracks (Martins and Palacios 2012) for high-mass
stars. We determined the stellar luminosity iteratively based on
the previously estimated $T_{\rm eff}=(33\,000\pm 1\,000)$ K and
$\log\,g=(3.85 \pm 0.15)$. As a result, we found the full set of
fundamental stellar parameters: the bolometric luminosity $\log
(L/L_{\odot})= 4.98 \pm 0.10$, the mass $M = (23 \pm
2)\ M_\odot$, and the radius $R = (9.5 \pm 1.5)\ R_\odot$.

\section*{DISCUSSION}
\noindent
One of the most important achievements of the INTEGRAL gamma-ray
observatory is the discovery of a hitherto unknown population of
Galactic X-ray binaries called Supergiant Fast X-ray
Transients. As has already been said, these are binaries
consisting of a neutron star with a strong magnetic field (an
X-ray pulsar) and an early-type supergiant. Only a few X-ray
binaries with supergiants were known before their
discovery. These were all quasi-persistent X-ray sources
emitting through accretion from the supergiant stellar wind, as
is expected for such binaries.  It was unclear why they are so
few. With the discovery of SFXTs the number of binaries with
supergiants has increased noticeably and already reaches
dozens. According to the SFXT model proposed by Grebenev and
Sunyaev (2007), the reason why persistently emitting binary
systems with supergiants are scarcely observed is the action of
a centrifugal barrier at the magnetospheric boundary of the
neutron star (the propeller effect, see Illarionov and Sunyaev
1975) halting the accretion, while the outbursts of SFXTs are
associated with the temporary overcoming of this barrier due to
a local increase in the density or a decrease in the velocity of
the stellar wind. Although qualitatively the model explains well
the observed phenomenon, a reliable determination of the orbital
and other parameters of the binary system for a number of SFXTs,
which is quite difficult to do, is needed to test it
quantitatively.

In this paper we presented the results of our long-term study of
the optical counterpart (companion star) to one of the canonical
SFXTs, IGR\,J17544-2619, at RTT-150. The orbital and physical
parameters of the binary system were determined reliably by the
methods of Doppler spectroscopy and by modeling its synthetic
spectra. The derived mass $M = (23 \pm 2) M_\odot$ and radius $R
= (9.5\pm1.5) R_{\odot}$ of the optical star differ from the
previously published values and correspond to an O9\,IV-V star
that ``sits'' more deeply in its Roche lobe. At the neutron star
mass $M_{X} = 1.4\ M_\odot$, according to Kepler's law, the
separation between the components is $a\simeq 35.5\ R_{\odot}$,
while the size of the Roche lobe of the optical counterpart,
$R_{r}=a\,[0.38+0.2\log(M/M_X)]\simeq22\ R_\odot$ (Paczynski
1971), is more than twice its actual radius
$R=(9.5\pm1.5)\ R_{\odot}$\footnote{Since the orbit is highly
  elliptical, the minimum distance between the components and
  the minimum radius of the Roche lobe are $\sim20.6 R_{\odot}$
  and $\sim12.8 R_{\odot}$, thus, at periastron the Roche radius
  exceeds the radius of the star by only $\sim30$\%. The rate of
  accretion from the wind at periastron is $\sim 3$ times higher
  than the average one in the orbit (see below).}. Contrary to the prevalent
opinion, the optical star of the binary does not belong
to supergiants. Nevertheless, it is close to them in its
many manifestations. In particular, the calculated
effective temperature $T_{\rm eff}=(33\,000 \pm 1\,000)$\,K is
sufficient for the formation of an intense stellar wind 
$\dot{M}_{W}\simeq(1-4)\times
10^{-7}\ M_{\odot}\ \mbox{yr}^{-1}$ (Vink et al. 2001;
van Buren 1985) capable of providing an accretion
rate onto the neutron star
$$ \dot{M}=\frac{1}{4}\left(\frac{r_a}{a}\right)^2\ \dot{M}_W \simeq$$
$$\simeq1.6\times 10^{-10}
\left(\frac{\dot{M}_W}{2\times10^{-7}\ M_{\odot}\ \mbox{yr}^{-1}}\right)\times$$
$$\times\left(\frac{v_W}{500\ \mbox{km s}^{-1}}\right)^{-4}\ M_{\odot}\
\mbox{yr}^{-1},$$
where $v_W$ is the presumed stellar wind velocity and
$r_a=2GM_X/v_W^2$ is the matter capture radius.  If the
gravitational energy of the accreting matter
$GM_X\dot{M}/R_X\simeq 0.2\dot{M}c^2$ is completely reprocessed
into radiation, then the X-ray luminosity of the binary can
reach $$L_X=1.8\times 10^{36}\
\left(\frac{\dot{M}_W}{2\times10^{-7}\ M_{\odot}\ \mbox{yr}^{-1}}\right)\times$$
$$\times\left(\frac{v_W}{500\ \mbox{km s}^{-1}}\right)^{-4}\ \mbox{erg s}^{-1}.$$

According to Grebenev and Sunyaev (2007), such a situation takes
place only when the spin period of the neutron star $P_s$
exceeds the so-called equilibrium period (Grebenev 2009)
$$P_s^*\simeq 0.98
\left(\frac{P_b}{5\ \mbox{days}}\right)^{37/21} \times $$
$$\times\left(\frac{B}{10^{12}\ \mbox{G}}\right)^{8/7}\left(\frac{v_W}{500\ \mbox{km s}^{-1}}\right)^{44/7} \times$$
$$\times\left(\frac{\dot{M}_W}{1\times10^{-5}\ M_{\odot}\ \mbox{yr}^{-1}}\right)^{-4/7}\ \mbox{s}.
$$
\begin{figure*}[th]
\centering
\includegraphics[width=150mm]{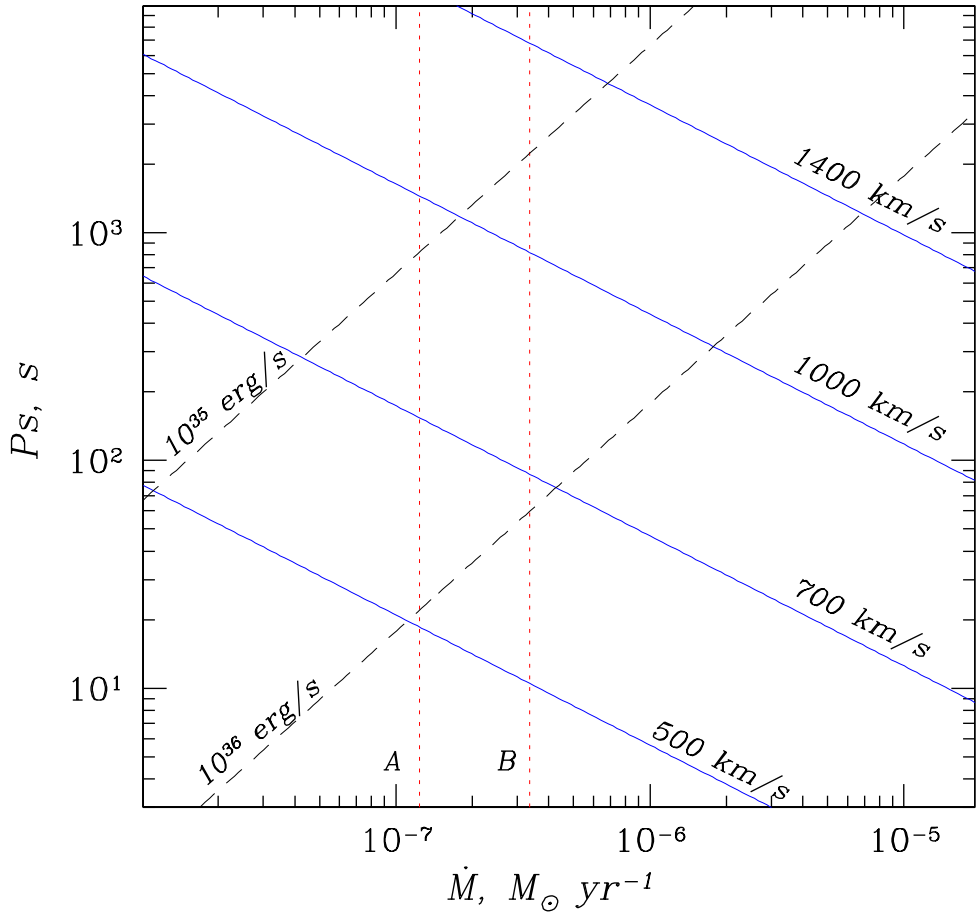}

\vspace{-4mm}

\caption{\rm Range of possible spin periods of the neutron star
  in IGR\,J17544-2619 as a function of the outflow rate of the
  stellar wind from the optical star. The solid (blue) lines
  indicate the equilibrium periods under different assumptions
  about the wind velocity. The dashed lines indicate the maximum
  achievable X-ray luminosities (in the absence of matter
  accumulation). The vertical (red) dotted lines indicate two estimates
  (see the text) of the wind outflow rate for a single OB star
  with the same parameters as those for the optical star of this
  binary system.}\label{psmdw}
\end{figure*}
\begin{figure*}[th]
\centering
\includegraphics[width=150mm]{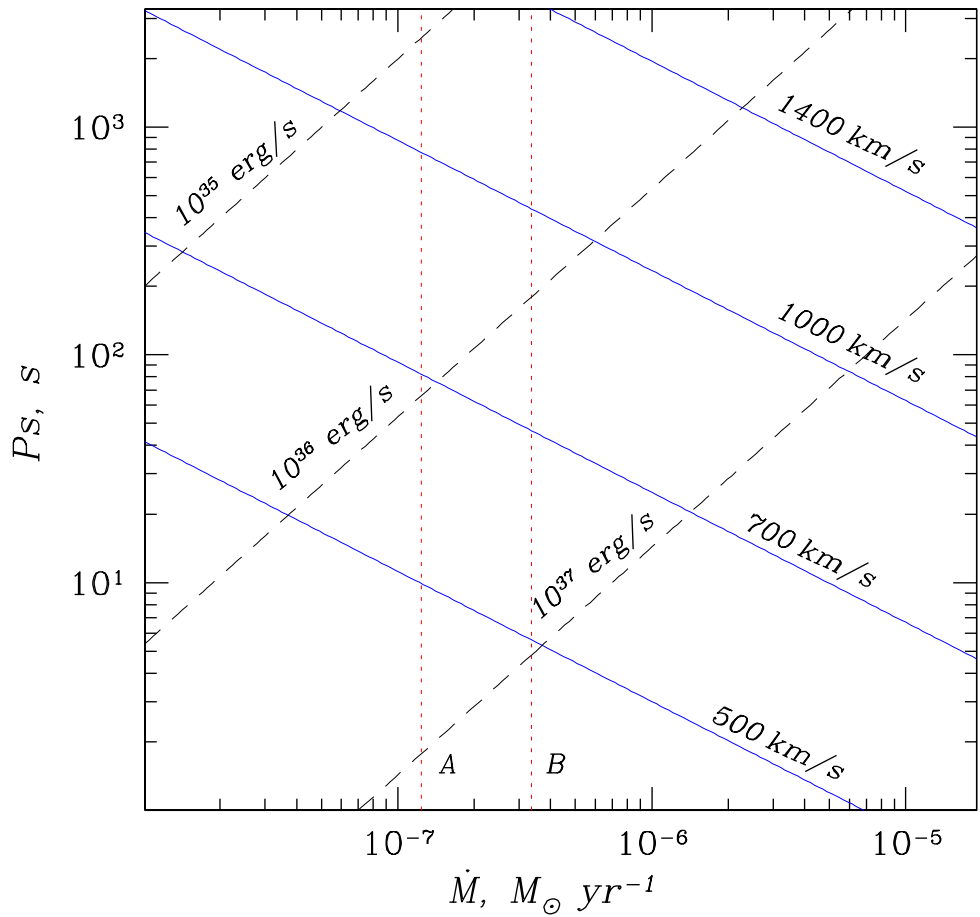}

\vspace{-4mm}

\caption{\rm The same as Fig.\,\ref{psmdw} but during the
  periastron passage by the neutron star. The rate of accretion 
  onto the neutron star and its luminosity are $\sim 3$ times
  higher than the average ones in the orbit (shown in
  Fig.\,\ref{psmdw}) at the same wind outflow rate. The spin
  periods shown by the blue lines are the equilibrium periods
  for the given accretion rate and may be slightly smaller than the
  actual period of the neutran star in
  IGR\,J17544-2619. }\label{psmdw2}
\end{figure*}

In Fig.\,\ref{psmdw} the solid lines indicate $P_s^*$ as a
function of the wind outflow rate $\dot{M}_{W}$ at the measured
parameters for IGR\,J17544-2619 and under different assumptions
about the wind velocity. It allows the range of values for the
as yet unknown period of the pulsar in this binary system to be
estimated. The vertical dotted lines in this figure indicate the
expected outflow rate of the stellar wind from a single OB star
with the same parameters as those for the optical component of
the system. Lines A and B correspond to the formulas from Vink
et al. (2001) and van Buren (1985), respectively. The dashed
lines indicate the maximum possible X-ray luminosities of the
source during outbursts (in the model by Grebenev and Sunyaev
(2007), in the absence of matter accumulation in the period
between outbursts). At a known observed luminosity these lines
bound the wind velocity. The figure shows that our estimates of
the parameters of the system and the optical star, on the whole,
agree well with the above SFXT model. The coincidence with the
model becomes even better during the periastron passage by the
neutron star in this system. Figure\,\ref{psmdw2} (similar to
Fig.\,\ref{psmdw}) shows the maximum possible X-ray luminosty of
the source and its spin period for this epoch under different
assumptions about the wind outflow rate and velocity. The
luminosity of about $10^{37}\ \mbox{erg s}^{-1}$ can be easily
reached here by the neutron star with a period  $\la10$ s.  At
the same time, the peak luminosity of $3\times
10^{38}\ \mbox{erg s}^{-1}$ recorded by Romano et al. (2015)
cannot be reached in the discussed range of wind outflow rates
that implies the ejection of more dense matter by the star at
that moment. We also see that the wind velocity cannot be very
high (cannot exceed greatly the parabolic velocity
$\sqrt{2GM/R}\sim1\,000\ \mbox{km s}^{-1}$). However, the wind
outflow rates estimated from the observations of single OB stars
(Vink et al. 2001; van Buren 1985) can be underestimated for the
star that is a component of an X-ray binary. According to the
observations by Bozzo et al. (2016), a weak (with $L_X\sim
10^{32}-10^{34}\ \mbox{erg s}^{-1}$) X-ray emission is recorded
from the neutron star of this system even in the period between
outbursts; it can heat up the surface of the optical star and
induce a stronger outflow of its  wind.

\section*{ACKNOWLEDGMENTS}
\noindent
This work was financially supported by the Russian Foundation
for Basic Research (project no. 16-32-50071), the Government of
Tatarstan (scientific project no. 15-42-02573), the
``Transitional and Explosive Processes in Astrophysics''
Subprogram of the Basic Research Program P-7 of the Presidium of
the Russian Academy of Sciences, and though subsidy
3.6714.2017\,/\,8.9 granted to the Kazan Federal University to
perform a State task in the area of scientific activity. We
thank TUBITAK, KFU, AST, and IKI for partial support in using
RTT-150 (the Russian-Turkish 1.5-m telescope in Antalya).


\vspace{2mm}

\begin{flushright}
{\it  Translated by V. Astakhov\/}
\end{flushright}
\end{document}